\documentclass[aps,physrev,preprint,groupedaddress]{revtex4-2}
\usepackage{graphicx}
\usepackage{caption}
\usepackage{subcaption}

\begin{document}

\title{Quarter-wave shift for an extra pass in an energy recovery linac}
\author{J. Michaud}
\affiliation{Université Paris-Saclay, CNRS/IN2P3, IJCLab, Orsay, France}
\author{A. Fomin}
\affiliation{Université Paris-Saclay, CNRS/IN2P3, IJCLab, Orsay, France}

\date{\today}

\begin{abstract}

Energy Recovery Linacs (ERLs) are advanced particle accelerators that recover the kinetic energy of the beam after interaction, significantly improving efficiency. In this paper we discuss an alternative operation mode for multi-pass ERLs with an extra pass at top energy, allowing a doubled beam current at the expense of an increased energy spread.

We explain the theoretical concept behind this approach, detailing how the phase shift enables an extra acceleration pass without requiring additional hardware. Preliminary simulations on the PERLE (Powerful ERL for Experiments) lattice explore the feasibility of this method. The results are discussed in the context of ERL performance optimization, and potential applications for future accelerator designs are explored.

\end{abstract}

\maketitle

\section*{Introduction}
\label{sec:intro}

Energy Recovery Linacs (ERLs) have emerged as a promising technology for high-power, high-efficiency particle acceleration. By reusing the beam's kinetic energy, ERLs achieve unprecedented energy efficiency while maintaining the beam quality from the injector to the experiment, making them ideal for applications such as free-electron lasers, nuclear physics experiments, high-energy particle collisions and electron based ion cooling.

However, the design of ERLs presents unique challenges, particularly in maximizing the circulating beam current while maintaining beam stability and energy recovery efficiency. Increasing the bunch charge to increase the current leads to an increase in collective effects strength and beam transportation along multiple passes can become extremely challenging.
\clearpage
In this paper, we propose an operation approach to achieve an extra pass in an ERL by introducing a {\it quarter-wave shift} in the radiofrequency (RF) phase, instead of the half-wave shift usually used in ERLs. This method allows the beam to pass through the linac without any net acceleration and enables an additional pass without the need for additional hardware. The methodology of this approach is discussed in Section~\ref{sec:concept}, followed by preliminary tests on the PERLE lattice in Section~\ref{sec:tests}. The implications and limitations of this method for ERL design and performance are explored in the Discussion (Section~\ref{sec:discussion}).

\section{Concept and methodology}
\label{sec:concept}

In an Energy Recovery Linac, particle bunches are typically accelerated near the crest of the RF voltage waveform, where the longitudinal electric field is at its maximum. Using the cosine representation of the RF voltage, $V_{acc} = V_0\cdot\cos(\phi)$, the bunch achieves maximum energy gain when its synchronous phase is $\phi_0 = 0$. After each acceleration pass, the bunch is bent through the spreader section into a return arc, whose length is an integer multiple of $\lambda_{\text{RF}}$. In a multi-pass ERL, this process is repeated across several linac passes and arcs to incrementally increase the beam energy.
Once the bunch reaches its target energy, it enters the high-energy return arc, which has a length of $\lambda_{\text{RF}}\cdot(N+1/2)$. This specific path length ensures that the bunch re-enters the linac with a phase shift of $\phi = \pi$ relative to the accelerating passes. Consequently, the RF field decelerates the bunch during subsequent passes, enabling efficient energy recovery.

\subsection{The direct consequences of a quarter-wave shift}

\begin{figure}
\includegraphics[width=0.7\textwidth]{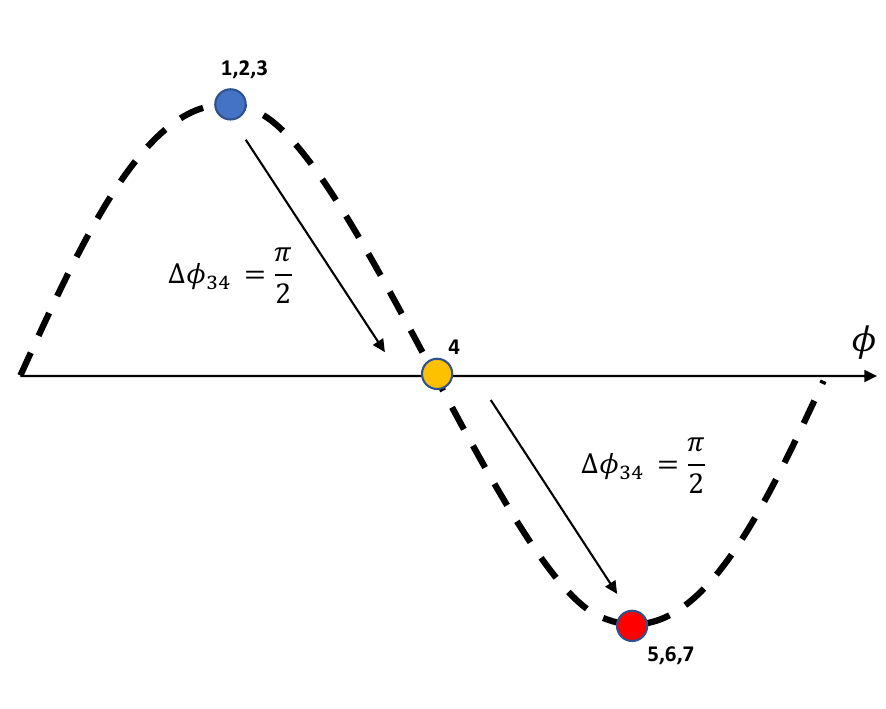}
\caption{
\label{phases_base} 
Schematic representation of the basic principle of the quarter-wave shift operation mode. The linac RF wave is represented by its phase and the effective acceleration voltage. The blue dot represents the on-crest accelerations, the red dot represents the off-crest decelerations occuring after the usual $\pi$ shift, while the yellow dot stands for the extra pass associated with the $\pi/2$ shift. Numbers from 1 to 7 stands for the pass number in the case of a 3x recirculating ERL.
}
\end{figure}

Rather than applying a $\lambda_{\text{RF}}/2$ path-length adjustment in the top-energy arc, the arc can be elongated by only $\lambda_{\text{RF}}/4$. In this configuration, the bunch re-enters the linac with a phase shift of $\pi/2$ instead of the standard $\pi$  shift. Since $\cos(\pi/2) = 0$, the bunch experiences no net acceleration and is bent back into the same arc for an additional pass. During this extra pass, it accumulates another $\pi/2$ shift, resulting in a total phase shift of $\pi$ relative to the linac RF. The bunch then follows its natural deceleration path toward the final dump. Figure~\ref{phases_base} illustrates this principle and serves as the baseline for the rest of this paper.

\paragraph*{Doubled beam current at interaction regions}

The direct effect of an extra pass at top energy is that interaction regions for experiments will experience a doubled beam current, with a direct impact on luminosity. However, because the same arc is used for recirculation, the machine tuning must be exceptionally precise to ensure that the beam properties are identical in both passes at the experimental area.

\paragraph*{Potential reduction of space charge forces}

Alternatively, instead of increasing the beam current, the bunch charge can be halved while maintaining the same current at the top-energy arc. This approach may be beneficial if the beam dynamics are space-charge dominated, as it could simplify transport through the machine by reducing space charge effects.

\paragraph*{Beam loading imbalance}

\begin{figure}
\includegraphics[width=0.7\textwidth]{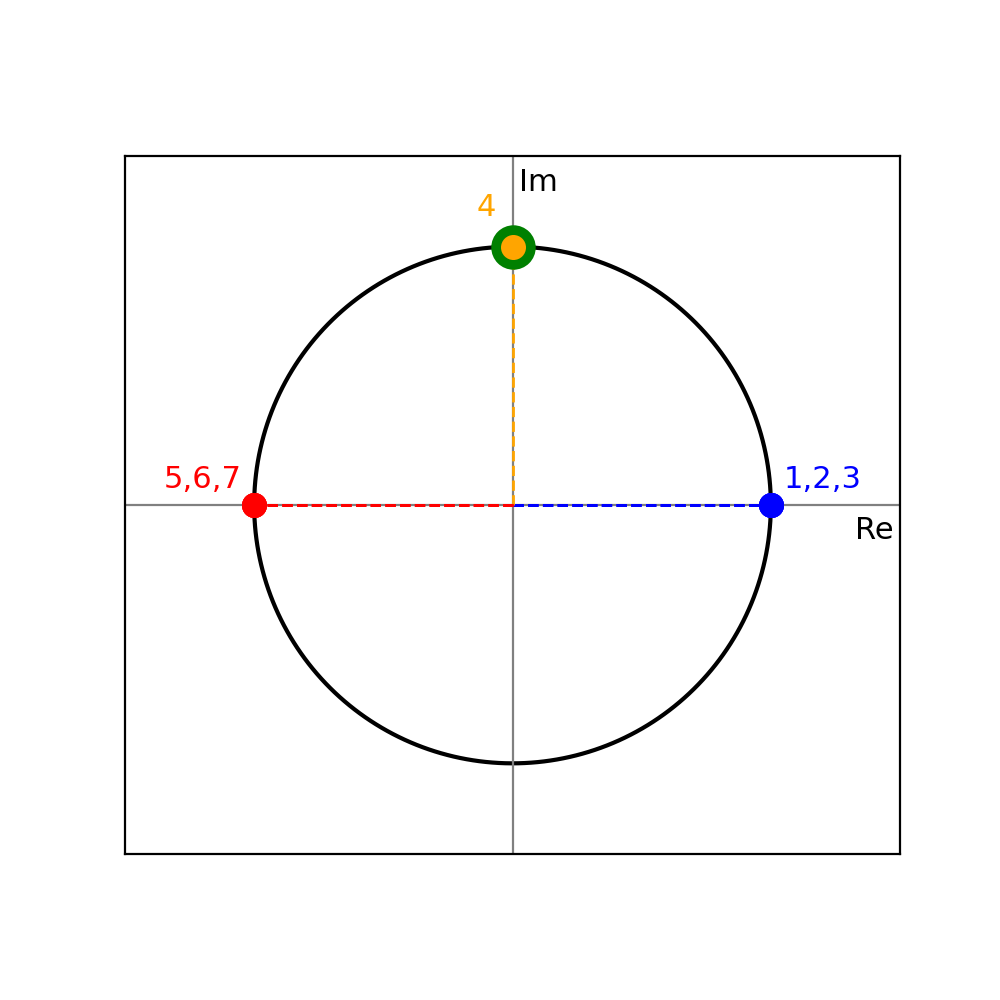}
\caption{
\label{beam_load_l4_base} 
Representation of RF beam loading in the complex plane. Blue dots represent the 3 acceleration passes, red dot the 3 deceleration passes, yellow dot the extra pass phase, and the green dot the resulting beam loading.
}
\end{figure}

The total beam loading induced by the many passes of the beam through the linac can be represented in the complex plane in a circle \cite{PWill}. In order to maintain energy recuperation, both the real and imaginary parts of the beam loading must be cancelled. The real part of beam loading can be nullified by using a symetric configuration of the phases, so that the energy gained and lost by the beams is at equilibrium. Figure \ref{beam_load_l4_base} shows that the extra pass will lead to the appearance of an imaginary beam loading component, inducing a strong inductive feedback in the cavity LLRF. 

\begin{figure}
\centering
\begin{subfigure}{.5\textwidth}
  \centering
  \includegraphics[width=1\linewidth]{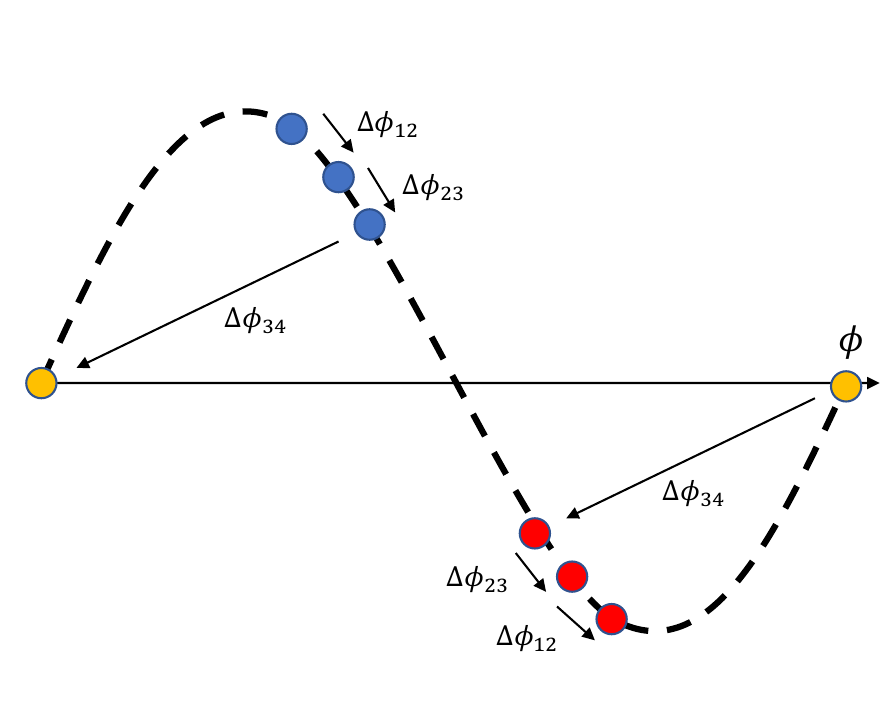}
  \caption{Phases location on RF potential}
  \label{phases_sol_pos}
\end{subfigure}%
\begin{subfigure}{.5\textwidth}
  \centering
  \includegraphics[width=1\linewidth]{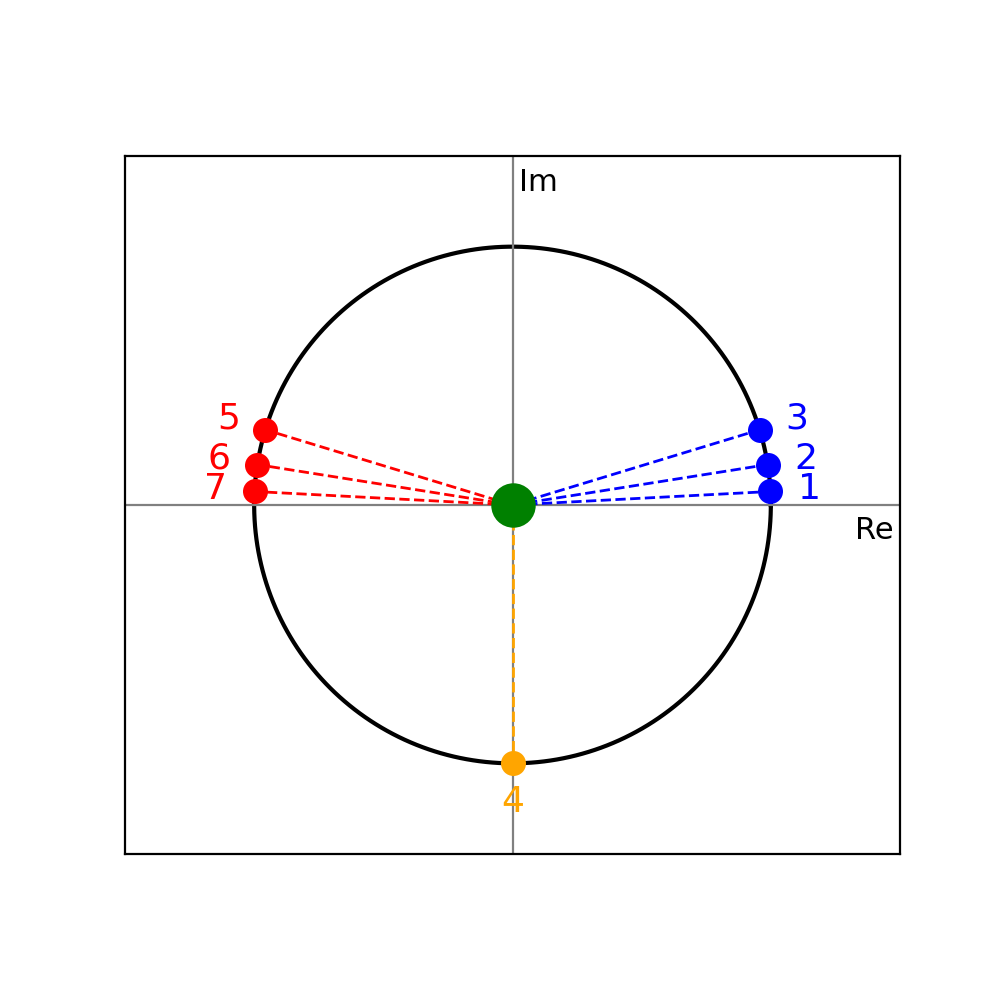}
  \caption{Beam loading}
  \label{beam_load_l4_sol_pos}
\end{subfigure}
\caption{Positive sign solution for the extra pass. Reprensentation of acceleration (blue) and deceleration (red) phases used. Yellow phase is for the extra pass, while green dot represents the total beam loading.}
\label{sol_pos}
\end{figure}

\begin{figure}
\centering
\begin{subfigure}{.5\textwidth}
  \centering
  \includegraphics[width=1\linewidth]{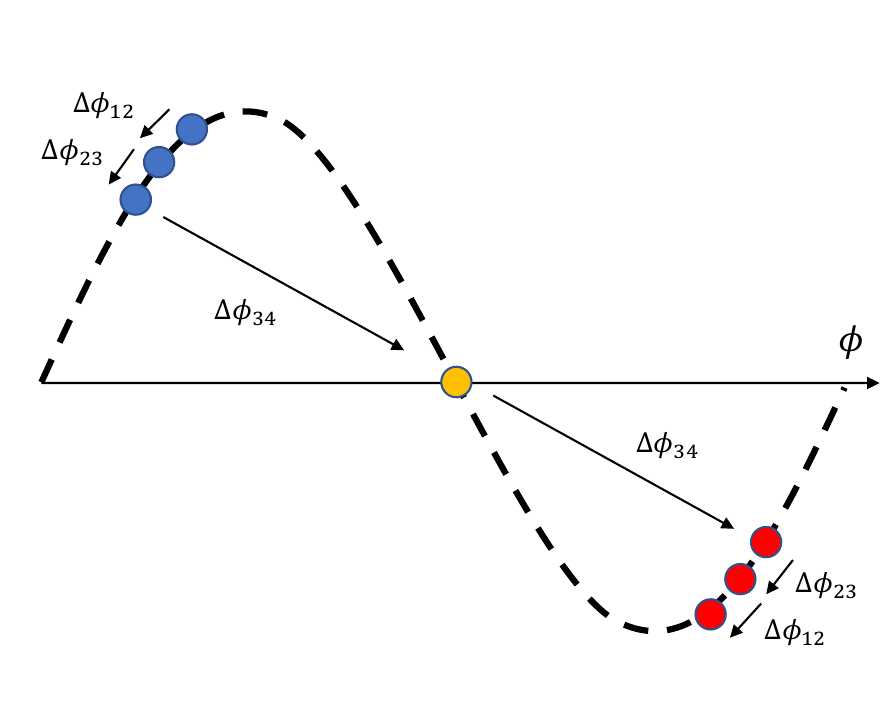}
  \caption{Phases location on RF potential}
  \label{phases_sol_neg}
\end{subfigure}%
\begin{subfigure}{.5\textwidth}
  \centering
  \includegraphics[width=1\linewidth]{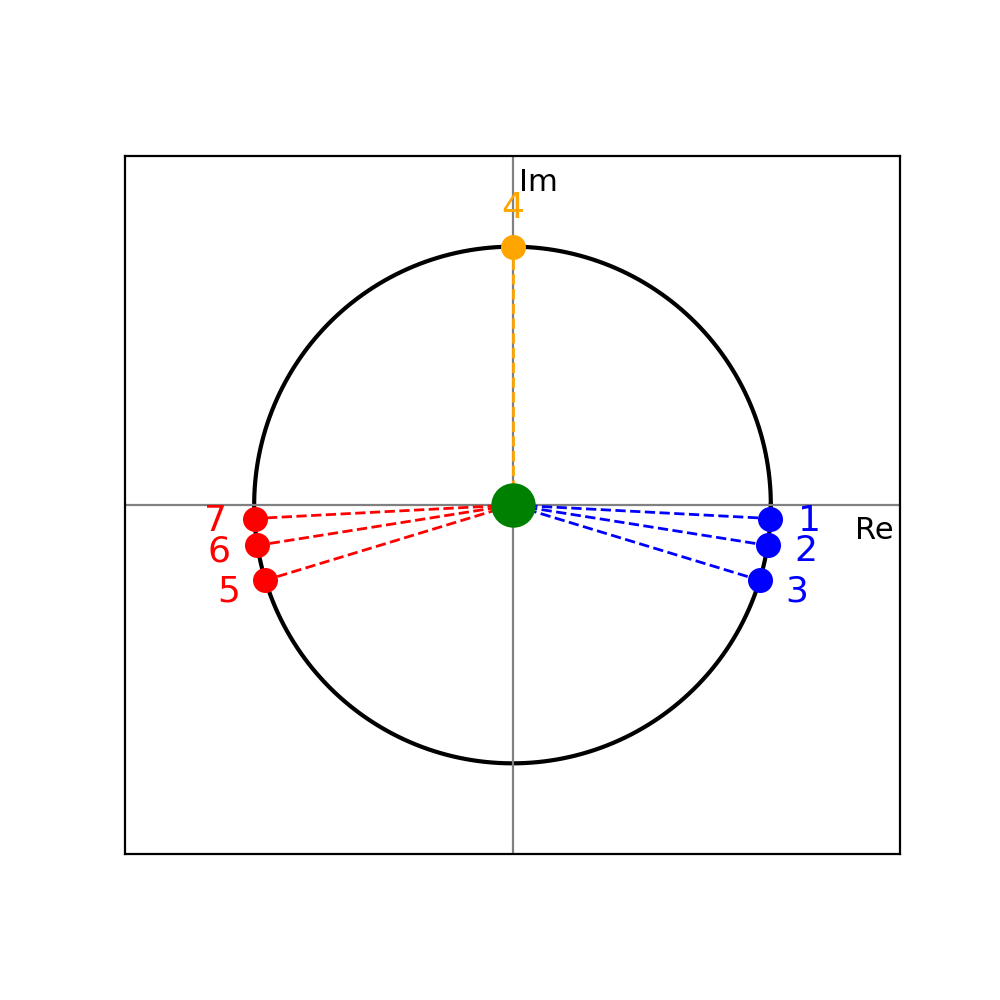}
  \caption{Beam loading}
  \label{beam_load_l4_sol_neg}
\end{subfigure}
\caption{Negative sign solution for the extra pass. Reprensentation of acceleration (blue) and deceleration (red) phases used. Yellow phase is for the extra pass, while green dot represents the total beam loading.}
\label{sol_neg}
\end{figure}

However, the resulting imaginary component can be cancelled by the use of off-crest acceleration and deceleration. To fully suppress the beam loading, the phases $\phi_i$ of the $i^{th}$ passes must satisfy the condition given in Eq.~\ref{condition_beam_loading}.

\begin{equation}
\label{condition_beam_loading}
    sin(\phi_1) + sin(\phi_2) + sin(\phi_3) = \pm\frac{1}{2}
\end{equation}

where the sign of the condition depends whether the phases are chosen to be on the rising or the falling side of the RF crest. This results in two possible solutions, that are represented in figures~\ref{sol_pos} and \ref{sol_neg} for the positive and negative sign cases, respectively.

\paragraph*{Energy spread growth}

The extra pass implies that the bunch enters the linac when the RF field provides no net acceleration, similar to a bunching cavity. The strong field gradient in this configuration leads to a sudden increase in the bunch energy spread, leading to potential losses (especially in dispersive areas) and a degradation of the experimental conditions.
This effect can be mitigate through three distinct approaches:
\begin{itemize}
    \item\textbf{Active energy spread cancellation} using longitudinal matching techniques. However, this usually demands multiple passes through a linac and therefore will be ineffective to provide a good beam quality to the experimental areas.
    \item \textbf{Cancellation of the RF-induced energy chirp} by applying an opposite chirp during the preceding passes.
    \item \textbf{Consistent reduction of the bunch length prior to the extra pass} to minimize the overlap between the bunch and the strong RF gradient region. The injection of a shorter bunch is possible, but injecting a too short bunch at low energy migth increase the sensitivity to CSR and space charge effects. Hence, bunch length compression must be performed via longitudinal matching. 
\end{itemize}

The two last methods can be performed simultaneously as they are inherently complementary when performing longitudinal matching. Additionnaly, theses approaches often require strong off-crest acceleration and phases on the same side of the crest which corresponds to the two solutions presented in figures \ref{sol_pos} and \ref{sol_neg}.

A compressive match of the bunch length can be achieved using the longitudinal dispersion term of the lattice, following the relation given by equation \ref{z_match}, where $z$ represents the longitudinal position of a particle relative to the center of the bunch, $R_{56}$ is the longitudinal dispersion of the lattice and $\delta = \Delta p/p_0$ is the relative momentum difference of the particle with respect to the reference particle (generally approximated as "relative energy difference" in relativistic cases). This relation impacts the individual positions of the particles within the bunch and will then cause a change in the total bunch length depending on the momentum spread ("energy spread").

\begin{equation}
\label{z_match}
    z = R_{56} \cdot \delta
\end{equation}

A longitudinal compression of the bunch can then be performed with a positive chirp (falling side of RF) and a negative $R_{56}$, or with a negative chirp (rising side of RF) and a positive $R_{56}$ value. 

Because the arcs in an energy recovery linac naturally exhibit negative longitudinal dispersion, the solution presented in Figure \ref{sol_pos} is generally prefered to relax the constrains on the machine quadrupoles.

\section{Experiments on a test lattice}
\label{sec:tests}
A proof-of-principle simulation was performed using a test lattice similar to the PERLE lattice, with the \textsc{BMAD} code~\cite{bmad}. An additional pass was added at the center of the machine to simulate the extra pass. This extra pass is a direct copy of the 3rd pass, except that the RF phase was set to $270^{\circ}$ instead of the $180^{\circ}$ necessary for the deceleration. This solution was chosen because it uses the natural longitudinal dispersion of the arcs.

For this first approach, we inserted taylor maps elements at the end of each arc to artificially insert longitudinal dispersion in the lattice, within the range actually achievable through proper matching. These elements are essentially unit transfer matrix with an additional term in the (5,6) position, allowing us to scan the potential longitudinal matching solutions without disrupting the Twiss parameters. $T_{566}$ values, which represents second order longitudinal dispersion can also be inserted to mitigate non-linear effects, and is normally tuned with the used of sextupoles.

Finally, as it was only a preliminary test, the phases were set directly in the linac cavities relatively to the synchronous phase without taking into account the effective path length of the machine.

\begin{figure}
\includegraphics[width=1\textwidth]{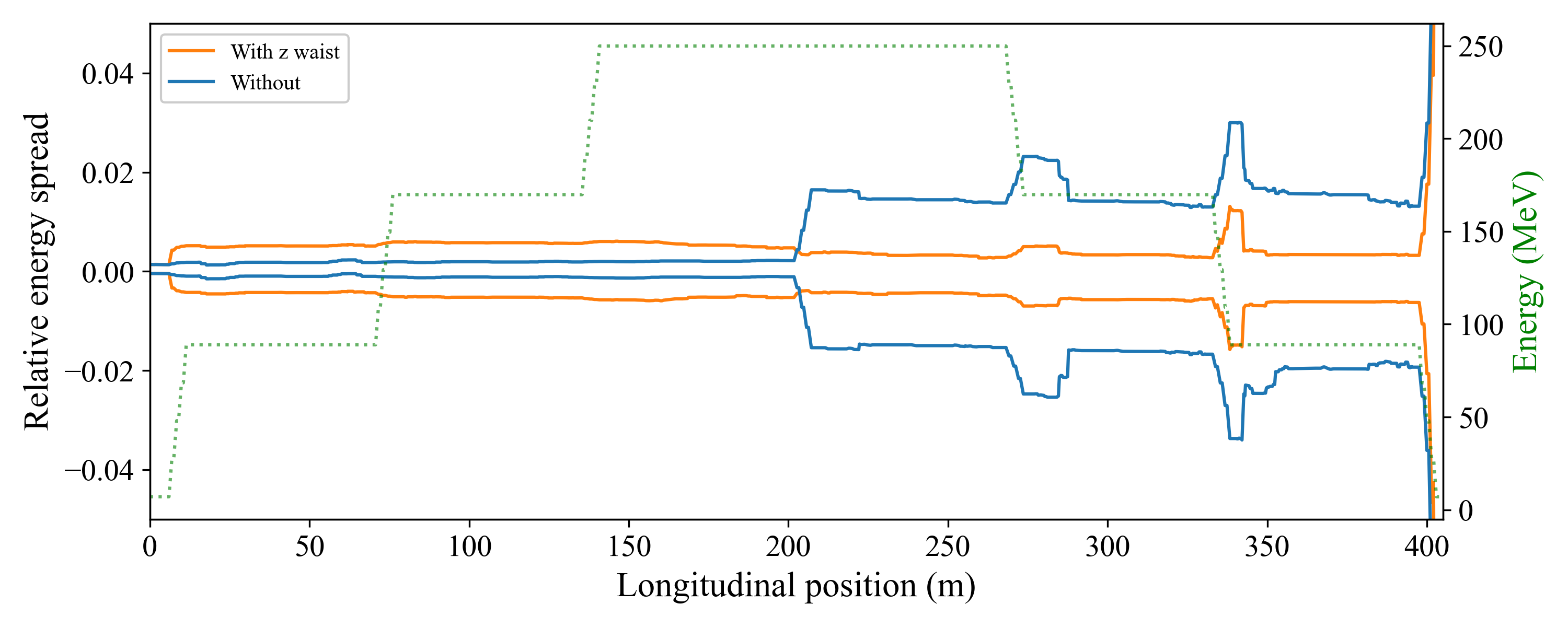}
\caption{
\label{pz_lambda4} 
Evolution of the energy spread across the test lattice, both with and without mitigation techniques.
}
\end{figure}

The first test, with no mitigation, showed an increase in energy spread from $0.12\%$ to up to $2\%$ immediately after the extra pass, as shown by the blue curve in figure \ref{pz_lambda4}, accompanied by a rise in particles losses from $0.005\%$ to approximately $60\%$.

To adress this, we performed an active compression of the bunch length during the first three passes by using the set of parameters given in table \ref{table_sol}. 
To minimize coherent synchrotron radiation (CSR) and space charge effects at injection, the initial bunch length was set to 3~mm, as this was previously shown to be the most stable configuration for PERLE\cite{bogacz}.

\begin{table}[h!]
\centering
\begin{tabular}{|c|c|c|c|}
\hline
$\phi_0$ & $\phi_1$ & $\phi_2$ & $\phi_3$ \\ \hline
$5$\textdegree & $10$\textdegree & $13$\textdegree & $270$\textdegree \\ \hline
\hline
$R_{56}$ merger &$R_{56}$ top arc 1 & $R_{56}$ mid arc 1 & $R_{56}$ bot arc 1 \\ \hline
-0.05 m & -0.15 m & -0.09 m & -0.07 \\ \hline
\hline 
$R_{566}$ bot arc 2 &$R_{56}$ top arc 2 & $R_{56}$ mid arc 2 & $R_{56}$ bot arc 2 \\ \hline
-7 m & -0.15 m & -0.12 m & -0.15 \\
\hline
\hline
\hline
$\sigma_z$ unmatched &$\sigma_z$ matched & $\delta$ unmatched & $\delta$ matched \\ \hline
3.03 mm & 1.04 mm & 0.016 & 0.004 \\\hline
\end{tabular}
\caption{List of parameters used for the optimisation on the PERLE lattice. Bunch length and energy spread for the unmatched and matched simulations are given on the last line.}
\label{table_sol}
\end{table}

\begin{figure}
\includegraphics[width=1\textwidth]{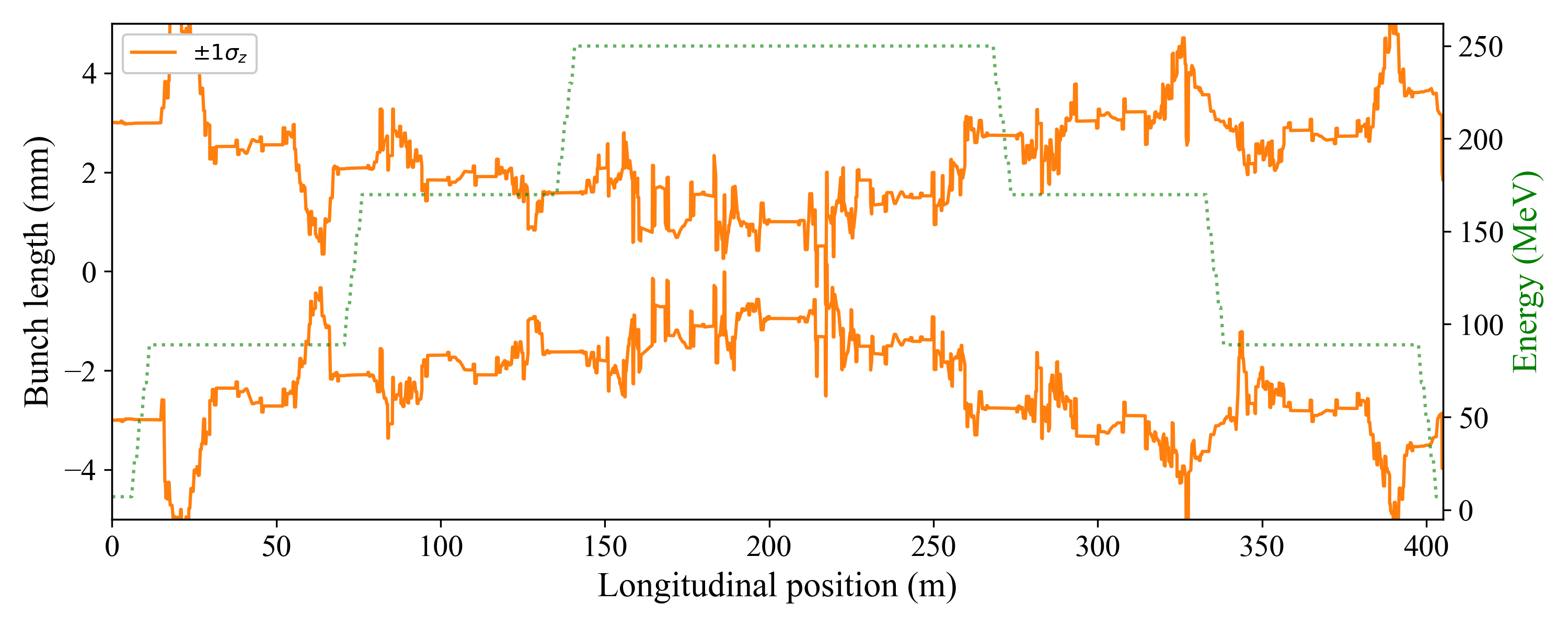}
\caption{
\label{z_lambda4} 
Bunch length evolution across the lattice, showing a waist formed by adjusting the Taylor elements and RF phases. This minimizes collective effects at low energy and reduces energy spread growth during the extra pass at high energy.
}
\end{figure}

The $R_{56}$ values of the Taylor elements and the RF phases were adjusted to create a "bunch length waist" across the full lattice. This ensures that the initial and final bunch lengths are maximized (to avoid collective effects at low energy) while the bunch length is minimized at top energy, as shown in Figure~\ref{z_lambda4}.

This resulted in a net reduction of the energy spread, as shown in Figure~\ref{pz_lambda4}, and reduced beam losses to below $1\%$. The energy spread in the experimental area was reduced from $0.016$ to $0.004$.

\section{Discussion}
\label{sec:discussion}
We presented a potential alternative operating mode for a multi-pass energy recovery linac, involving an extra pass at high energy. This additional pass can either double the beam current at the experimental areas or maintain the same current with a halved bunch charge.

Preliminary tests revealed a significant increase in both energy spread and beam losses. Our initial mitigation tests on the PERLE lattice were insufficient to fully suppress beam losses or maintain the initial energy spread. However, a more thorough optimization of the lattice, or the design of a lattice specifically tailored for this mode, could yield successful results.

The extra pass could be particularly beneficial for experiments that do not require a small energy spread.

\bibliography{biblio}

@article{PWill,
  title = {Construction of self-consistent longitudinal matches in multipass energy recovery linacs},
  author = {P\'erez Segurana, Gustavo and Bailey, Ian R. and Williams, Peter H.},
  journal = {Phys. Rev. Accel. Beams},
  volume = {25},
  issue = {2},
  pages = {021003},
  numpages = {13},
  year = {2022},
  month = {Feb},
  publisher = {American Physical Society},
  doi = {10.1103/PhysRevAccelBeams.25.021003},
  url = {https://link.aps.org/doi/10.1103/PhysRevAccelBeams.25.021003}
}

@article{bmad,
  author     = "D. Sagan",
  title      = "{Bmad: A relativistic charged particle simulation library}",
  booktitle  = "{Computational accelerator physics. Proceedings, 8th
                International Conference, ICAP 2004, St. Petersburg,
                Russia, June 29-July 2, 2004}",
  journal    = "Nucl. Instrum. Meth.",
  volume     = "A558",
  number     = "1",
  year       = "2006",
  pages      = "356-359",
  issn       = "0168-9002",
  doi        = "https://doi.org/10.1016/j.nima.2005.11.001",
  note       = "Proceedings of the 8th International Computational Accelerator Physics Conference",
  SLACcitation = "%%CITATION = NUIMA,A558,356;%%"
}

@article{bogacz,
  title = {Beam dynamics driven design of powerful energy recovery linac for experiments},
  author = {Bogacz, S. A. and Andr\'e, K. D. J. and Br\"uning, O. and Holzer, B. J. and Hounsell, B. R. and Klein, M. and Militsyn, B. L. and Williams, P. H. and Segurana, G. P\'erez and Bailey, I. and Apsimon, R. and Setiniyaz, S. and Abukeshek, R. and Barbagallo, C. and Abdillah, M. Ben and Bruni, C. and Duchesne, P. and Duthil, P. and Fomin, A. and Guyot, C. and Kaabi, W. and Michaud, J. and Olry, G. and Perrot, L. and Reynet, D. and Roux, R. and Stocchi, A. and Wurth, S. and Abualrob, H. and Baylac, M. and Bouly, F. and Jacquot, B.},
  journal = {Phys. Rev. Accel. Beams},
  volume = {27},
  issue = {3},
  pages = {031603},
  numpages = {15},
  year = {2024},
  month = {Mar},
  publisher = {American Physical Society},
  doi = {10.1103/PhysRevAccelBeams.27.031603},
  url = {https://link.aps.org/doi/10.1103/PhysRevAccelBeams.27.031603}
}

\end{document}